\documentclass{article}
\usepackage{spconf,amsmath,graphicx}
\usepackage{booktabs}  %  引入三线表宏包
\usepackage{amsthm,amsmath,amssymb}
\usepackage{xcolor}
\usepackage{tablefootnote}
\usepackage{hyperref}
\usepackage{multirow}
\usepackage[normalem]{ulem}

\title{MMSpeech: Multi-modal Multi-task Encoder-Decoder Pre-training for speech recognition}
\name{Xiaohuan Zhou$^*$, Jiaming Wang$^*$, Zeyu Cui, Shiliang Zhang, Zhijie Yan, Jingren Zhou, Chang Zhou\textsuperscript{$\dagger$}
\thanks{* Equal contribution.}
\thanks{$\dagger$ Corresponding author}
}
\address{DAMO Academy, Alibaba Group, China \\
\{shiyi.zxh, wangjiaming.wjm, zeyu.czy, sly.zsl, zhijie.yzj, jingren.zhou, ericzhou.zc\}@alibaba-inc.com}
\begin{document}

\maketitle
\begin{abstract}
In this paper, we propose a novel multi-modal multi-task encoder-decoder pre-training framework~(MMSpeech) for Mandarin automatic speech recognition~(ASR), which employs both unlabeled speech and text data. The main difficulty in speech-text joint pre-training comes from the significant difference between speech and text modalities, especially for Mandarin speech and text. 
Unlike English and other languages with an alphabetic writing system, Mandarin uses an ideographic writing system where character and sound are not tightly mapped to one another.
Therefore, we propose to introduce the phoneme modality into pre-training, which can help capture modality-invariant information between Mandarin speech and text. 
Specifically, we employ a multi-task learning framework including five self-supervised and supervised tasks with speech and text data. 
For end-to-end pre-training, we introduce self-supervised speech-to-pseudo-codes~(S2C) and phoneme-to-text~(P2T) tasks utilizing unlabeled speech and text data, where speech-pseudo-codes pairs and phoneme-text pairs are a supplement to the supervised speech-text pairs. To train the encoder to learn better speech representation, we introduce self-supervised masked speech prediction~(MSP) and supervised phoneme prediction~(PP) tasks to learn to map speech into phonemes. Besides, we directly add the downstream supervised speech-to-text~(S2T) task into the pre-training process, which can further improve the pre-training performance and achieve better recognition results even without fine-tuning.
Experiments on AISHELL-1 show that our proposed method achieves state-of-the-art performance, with a more than 40\% relative improvement compared with other pre-training methods.
\end{abstract}
\begin{keywords}
ASR, pre-training, encoder-decoder
\end{keywords}
\section{Introduction}
\label{sec:intro}
% Recent end-to-end ASR models\cite{graves2012sequence,graves2014towards,hannun2014deep,chan2015listen} follow two main approaches, Connectionist Temporal Classification (CTC) \cite{graves2006connectionist}  and encoder-decoder models \cite{bahdanau2014neural}. The CTC model considers only the acoustic information for recognition, while the encoder-decoder model\cite{chan2015listen} considers both acoustic features and language knowledge. Specifically, the encoder is responsible for acoustic modeling (AM), and the decoder is used for language modeling (LM). A cross-attention mechanism (CA) is to build a relationship between AM and LM. Furthermore, the Mandarin speech recognition task benefits more from language modeling than other phonetic languages, such as English, since Chinese is an ideographic language. Therefore, the encoder-decoder model usually outperforms the CTC model.
Recently, research on pre-training has been widely investigated and significantly improves the performance of downstream speech tasks, such as automatic speech recognition~(ASR). 
In general, pre-training methods for ASR can be roughly divided into two branches, namely encoder pre-training and encoder-decoder pre-training methods. For encoder pre-training methods, a large amount of unlabeled speech data is used to help the encoder learn the ability to extract the universal speech representation~\cite{schneider2019wav2vec,baevski2020wav2vec,hsu2021hubert,baevski2022data2vec}. 
For example,
%wav2vec~\cite{schneider2019wav2vec} employs a multi-layer convolutional neural network as the backbone and is optimized by a noise contrastive binary classification task. HuBERT~\cite{hsu2021hubert} utilizes an offline clustering method to generate aligned target labels, and then a predictive loss is applied to pre-train the encoder. 
% Data2Vec~\cite{baevski2022data2vec} is trained by generating representations using the teacher encoder based on the full input and then regressed by the student encoder of the same architecture based on a masked version of the input. 
As only the encoder is pre-trained, they usually employ connectionist temporal classification~(CTC)~\cite{graves2006connectionist} based models for downstream ASR tasks. 
Since encoder-decoder based ASR models~\cite{chan2015listen,mohamed2019transformers} usually obtain better recognition performance, encoder-decoder pre-training methods are further proposed.
% To fine-tune encoder-decoder based ASR models~\cite{chan2015listen}, which can usually obtain better recognition performance, encoder-decoder pre-training methods are further proposed. 
% \czy{Encoder-decoder methods, which replace the CTC layer with more complex decoder layer, usually obtain better recognition performance. To get a downstream encoder-decoder ASR model,  encoder-decoder pre-training methods are further proposed. }
These methods are usually optimized within a multi-task learning framework.
For example, Speech2C~\cite{ao2022pre} introduces two tasks using speech-only data via pseudo codes. One is to predict the pseudo codes via masked language modeling, and the other is to learn the reconstruction of pseudo codes. 
Recently, multi-modal pre-training has achieved great success in both cross-modal and single-modal downstream tasks.
% OFA~\cite{wang2022unifying} proposes a multi-modal multi-task pre-training framework via a unified I/O and task representation, showing superior performance on a variety of downstream vision-language tasks. 
% As for speech processing, 
SpeechT5~\cite{ao2021speecht5} leverages unlabeled speech and text data to improve the speech-text cross-modal tasks. Besides, STPT~\cite{tang2022unified} takes advantage of the supervised speech data to improve the multi-modal pre-training. However, there are still some aspects needed to be further investigated: (1) Existing encoder-decoder pre-training works are mainly exclusively for English while almost none for Mandarin; (2) Unlabeled text data is underestimated and less explored in speech pre-training literature; (3) The complementarity between tasks of different works is not fully exploited.

To address these problems, we introduce the phoneme modality into pre-training, which can capture modality-invariant information between speech and text; that is, both speech and text can be uniquely mapped to a phoneme sequence. It is especially beneficial for Mandarin ASR pre-training since Mandarin is an ideographic language. Unlike English and other languages with an alphabetic writing system, Mandarin uses an ideographic writing system where character and sound are not tightly mapped to one another~\cite{zhou2015homophone}. There is an enormous difference between Mandarin speech and text, while phonemes bridge the gap.
Besides, we introduce 292G text data from M6-Corpus~\cite{lin2021m6} for pre-training, which is much larger than previous works~\cite{ao2021speecht5,Tang2021AGM,tang2022unified} where only 1.8G text data are used. Experiments demonstrate that enough unlabeled text data is also useful like unlabeled speech data. 
To fully exploit the value of speech and text data, we employ five self-supervised and supervised tasks for multi-task pre-training utilizing speech and text data, including phoneme-to-text (P2T), speech-to-pseudo-codes (S2C), masked speech prediction (MSP),  phoneme prediction (PP), and supervised speech-to-text (S2T) tasks.
% P2T gives the model better language generation ability by unlabeled text data. MSP, PP, and S2C make the model understand audio better. S2T is the common ASR task.

In this way, we propose a novel multi-modal multi-task encoder-decoder pre-training framework~(MMSpeech) for Mandarin ASR. Five tasks are employed for multi-task pre-training. 
To utilize text data, we introduce the P2T task, which is a modified version of text-infilling~\cite{lewis2019bart,ao2021speecht5}. We apply phonemes rather than Chinese characters as the input, which can effectively reduce the difference between Mandarin speech and text and ease them to share an encoder.
% To utilize text data, we introduce the P2T task with unlabeled text data.
% Note that P2T is a modified version of text-infilling~\cite{lewis2019bart,ao2021speecht5} where we apply phonemes rather than Chinese characters as the input, which can effectively reduce the difference between Mandarin speech and text and ease them to share an encoder.
Experiments prove that the improvements achieved by the P2T task pre-training can not be replaced by an external language model~(LM), demonstrating that P2T not only learns the grammatical rules of text but also learns the connection between pronunciation information and text.
As for speech data, we introduce a self-supervised S2C task that translates speech to pseudo-codes within a sequence-to-sequence manner. Besides, to train the encoder to extract better speech representation for ASR, an MSP task is introduced for the encoder, which predicts phoneme distributions by masked language modeling with unlabeled speech data. We choose to predict phoneme distributions rather than hidden states like ~\cite{baevski2020wav2vec,hsu2021hubert,baevski2022data2vec} since phonemes containing only pronunciation information are a bridge between speech and text. Here, we also introduce a supervised PP task to assist MSP, which also predicts phonemes but with paired speech-text data and a CTC loss.
Furthermore, introducing the downstream S2T task can further improve the pre-training performance and directly obtain excellent recognition results even without fine-tuning. 
% Another advantage of S2T is that we can directly estimate the quality of pre-training in the pre-training stage according to the results of S2T, which can facilitate the confirmation of pre-training-related configuration. 
% 
% \czy{MMSpeech\footnote{code will be released when the paper is released.} is realized based on OFA~\cite{wang2022unifying}, a 
% multi-modal multi-task pre-training framework via a unified I/O and task representation.  }
Experiments on AISHELL-1 corpus show that our proposed method achieves the state-of-the-art~(SOTA) performance, with a more than 40\% relative improvement compared with other pre-training methods.

\section{Related Work}

\noindent
\textbf{Speech Self-supervised Learning:} 
CPC~\cite{oord2018representation} and Wav2Vec ~\cite{schneider2019wav2vec} are two early works that propose to utilize contrastive loss for self-supervised speech representation learning. Wav2Vec 2.0~\cite{baevski2020wav2vec} applying BERT~\cite{devlin2018bert} pre-training on speech is the first self-supervised model to outperform purely supervised methods on ASR. Specifically, it masks the speech input in the latent space and solves a contrastive task distinguishing the quantized version of the correct features from several negative samples. A series of BERT pre-training methods for speech self-supervised learning is then proposed, which adopt different targets for the masked frames.
HuBERT~\cite{hsu2021hubert} utilizes an offline clustering method to generate aligned target labels, and then the model is trained to predict the cluster indices of the masked frames. Based on HuBERT, WavLM~\cite{chen2022wavlm} uses a larger dataset and data augmentation during pre-training.
Data2Vec~\cite{baevski2022data2vec} predict latent representations of the full input data based on a masked version of the input in a self-distillation setup. Thanks to the continuous and contextualized target, Data2Vec achieves improvements. With a pre-trained encoder only, CTC models usually outperform the encoder-decoder model in ASR. Speech2C~\cite{ao2022pre} and Wav2Seq~\cite{wu2022wav2seq} are further proposed to boost the performance of the encoder-decoder. They utilize speech-only data and apply speech-pseudo-codes pairs for sequence-to-sequence manner training where the pseudo codes are generated by an existing pre-trained encoder.

\noindent
\textbf{Multi-modal Pre-training:}
Multi-modal pre-training has achieved great success in computer vision~\cite{lu2019vilbert,radford2021learning,wang2022unifying} and speech processing~\cite{bapna2021slam,ao2021speecht5,tang2022unified} area. A series of research focuses on leveraging text data to improve language modeling ability in speech-to-text tasks~\cite{han2021learning,Tang2021IST,Tang2021AGM}, where the speech and text inputs share an encoder and use the decoder to generate text sequences. To alleviate the difficulty of learning representations of speech and text within a shared encoder, SLAM~\cite{bapna2021slam} and mSLAM~\cite{bapna2022mslam} propose to leverage extra supervised speech-text tasks to align the representations of speech and text. Both unlabeled speech and text data are used in these works; however, only an encoder is pre-trained.

Since the encoder-decoder based models is a more common choice for speech processing tasks, SpeechT5~\cite{ao2021speecht5} and STPT~\cite{tang2022unified} are proposed for encoder-decoder pre-training with unlabeled text and speech data, which are most related to our method. SpeechT5~\cite{ao2021speecht5} proposes to unify all the spoken language processing tasks as a speech/text to speech/text problem via an encoder-decoder framework. The differences between our method and SpeechT5 are: (1) MMSpeech is mainly designed for the ASR task, while some designs of SpeechT5 are not considered for the ASR; for example, SpeechT5 utilizes the speech reconstruction task for pre-training, whose loss follows the text-to-speech (TTS) task and is of little help for the ASR task. (2) SpeechT5 proposes a cross-modal vector quantization method to capture the modality-invariant information between speech and text, while we propose that the phonemes are a natural bridge between speech and text. 

Another related work is STPT~\cite{tang2022unified}, which proposes a multi-task learning framework including four self-supervised and supervised tasks and achieves improvements on downstream ASR and speech translation tasks. The differences between our method and STPT are: (1) During pre-training, the decoder of STPT is only exposed to the text data while never exposed to the speech data. MMSpeech proposes to use the unlabeled speech to generate speech-pseudo-codes pairs for encoder-decoder pre-training, which enhance the abilities of decoder to locate and encapsulate speech information. (2) MMSpeech conducts experiments on the Mandarin dataset, while STPT is for English. Mandarin is an ideographic language and has a much larger number of homophones than English. Therefore, we observe some different experimental conclusions from STPT.

\begin{figure*}
\centering
\includegraphics[width=0.8\linewidth]{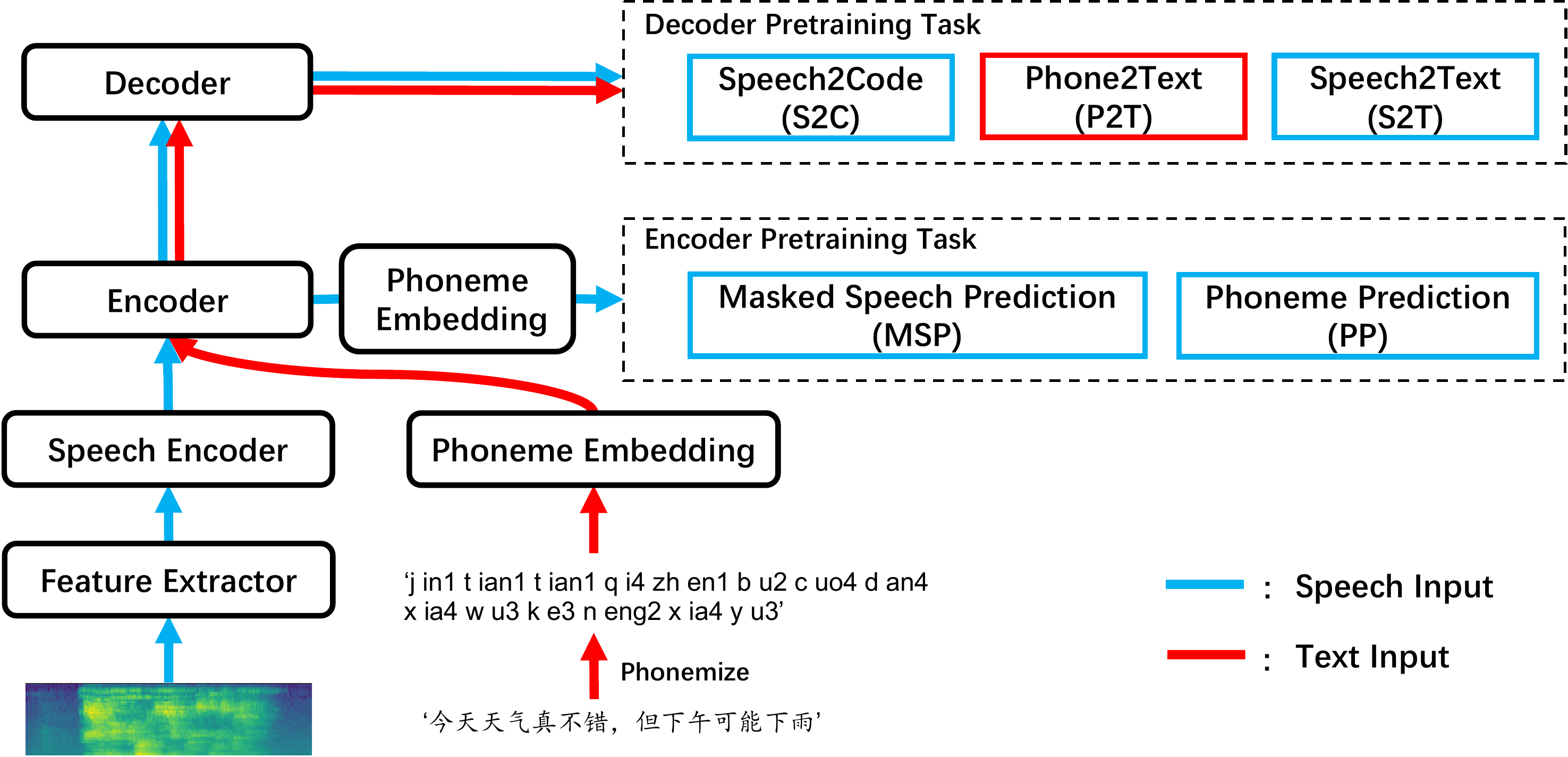}
% \caption{MMSpeech. The modules S2T are used in the final ASR model and the others are used to auxillate the pre-training. The blue, orange, purple, green, yellow lines represent different tasks' data flow in the model for the \textcolor{blue}{MSP}, \textcolor{orange}{PP}, \textcolor{purple}{S2C}, \textcolor{green}{P2T} and \textcolor{yellow}{S2T} tasks. \zc{Need a clearer Figure.}}
\caption{MMSpeech. The modules S2T are used in the final ASR model and the others are used to auxillate the pre-training. The blue, red lines represent speech and text  data flow in the model.}
\label{model}
\end{figure*}

\section{Methods}
In this section, we elaborate the proposed MMSpeech based on encoder-decoder framework. As illustrated in Fig.~\ref{model}, MMSpeech consists of five tasks and apply speech and text data as the input. 
% Specifically, self-supervised MSP and supervised PP tasks are introduced for encoder pre-training while supervised P2T, P2C and S2T tasks are introduced for decoder pre-training.

\subsection{Model Architecture}
As shown in Fig.~\ref{model}, MMSpeech mainly employs the encoder-decoder architecture. The encoder network consists of a speech feature extractor, a speech encoder and a shared encoder. Specifically, the speech feature extractor is a multi-layer convolutional network while the speech encoder and the shared encoder are multiple transformer layers with multi-head self attention~\cite{vaswani2017attention}. The decoder network is also multiple transformer layers, which is similar to the encoder except for masked self-attention and cross-attention.

\subsection{Encoder-Decoder Pre-Training Tasks}
\label{sec:dec-pre}
For encoder-decoder pre-training, we introduce self-supervised phoneme-to-text~(P2T) and speech-to-pseudo-code~(S2C) tasks which utilize unlabeled text data and unlabeled speech data.

\subsubsection{Phoneme-to-text}
We propose to use the P2T task utilizing large-scale unlabeled text data for pre-training, which is a modified version of text-infilling. 
Specifically, we convert input Mandarin texts into phoneme sequences via a mapping dictionary such as the open-sourced ``pypinyin" python package. Besides, we add noise to the phoneme sequences by masking or replacing token spans. Then the phoneme embedding $\mathbf{E}=(\mathbf{e}_1,...,\mathbf{e}_I)$ and shared encoder is employed to extract phoneme features $\mathbf{H}^{e}$. 
The decoder reconstructs the text sequence $\mathbf{Y}^t=\left(\mathbf{y}^{t}_{1}, \ldots, \mathbf{y}^{t}_{L} \right)$ like the text-filling task based on $\mathbf{H}^{e}$. The task is optimized by maximizing cross-entropy:
\begin{equation}
\mathcal{L}_{P2T} = - \sum_{l=1}^{L} log p(\mathbf{y}^t_l|\mathbf{y}^t_{l-1},\mathbf{H}^e)
\end{equation}

The introduction of phonemes here brings two advantages: (1) It is easy for phoneme and speech inputs to share one encoder. Specifically, Mandarin is an ideographic language where character and sound are not tightly mapped to one another, and there is a large difference between Mandarin speech and text. Phoneme closes the gap between speech and text since it represents the corresponding pronunciation of the text and is a closer form to speech~\cite{Tang2021AGM,tang2022unified}. (2) The phonemes-text pairs are a high-quality supplement to the supervised speech-text pairs for ASR. 
P2T is too simple for English, a phonetic language; for example, 76\% of English words in the Carnegie Mellon Pronouncing Dictionary can be recovered from a phoneme sequence deterministically~\cite{Tang2021AGM}.
However, compared to English, Mandarin has a much larger number of homophones~\cite{zhou2015homophone}, and P2T is no longer a simple task for Mandarin. As shown in Section \ref{sec:exp}, even with an external language model, P2T can still bring significant improvements, which proves that P2T learns not only the language modeling abilities but also the relationship between speech and text.
Note that we share the phoneme embedding $E$ here with the PP task as described in Section~\ref{sec:PP}, which helps align the phoneme representation and speech representation.

\subsubsection{Speech-to-pseudo-code}
For the S2C task, we automatically annotate the unlabeled speech with pseudo-codes, then train the encoder-decoder with the speech-pseudo-codes pairs. We introduce a reconstruction loss following Speech2C~\cite{ao2022pre}, which generates pseudo-codes $\mathbf{Y}^c=\left(\mathbf{y}^{c}_{1}, \ldots, \mathbf{y}^{c}_{L} \right)$ in an autoregressive fashion:
\begin{equation}
    \mathcal{L}_{\text{S2C}} = - \sum_{l=1}^{L} \text{log} p(\mathbf{y}^c_l|\mathbf{y}^c_{l-1},\mathbf{H})
\end{equation}
where $\mathbf{H}$ is the encoder output extracted from speech input $\mathbf{X}$\footnote{In this paper, the speech input is the log Mel-filterbank feature rather than the raw audio waveform~\cite{baevski2020wav2vec,hsu2021hubert,baevski2022data2vec}}. Now we describe how to obtain pseudo-codes for the unlabeled speech. Specifically, a sequence of hidden states is first extracted from speech using a pre-trained encoder such as Data2Vec model~\cite{baevski2022data2vec}. Then K-means clustering is applied to discretize these hidden states and obtain a sequence of hidden units~\cite{hsu2021hubert}. To shorten the sequence length like~\cite{wu2022wav2seq}, we deduplicate these units~(e.g., ``1 1 1 2 3 3" will be processed to ``1 2 3" ) 
and further adopt a byte-pair encoding~(BPE)~\cite{gage1994new} model to generate the final pseudo-codes. 
% Compared to the hidden units, the generated pseudo-codes are more relevant to the semantics of speech. 
% Compared to the hidden units, pseudo-code sequences are closer to text sequences in sequence length and characteristics of speech velocity agnostic.
Compared to the hidden units, the generated pseudo-codes are closer to text tokens in the amount of information. 
Therefore, the speech-pseudo-codes pairs are also a high-quality supplement to the supervised speech-text pairs for sequence-to-sequence manner training. The abilities of decoder to locate and encapsulate speech information needed to generate the next token are enhanced.
% The S2C task is actually similar with ASR, with pseudo-codes instead real texts and thus can also help the decoder learn how to generate target sequences.

\subsection{Encoder Pre-Training Tasks}
For encoder pre-training, we introduce self-supervised masked speech prediction~(MSP) and supervised phoneme prediction~(PP) tasks which utilize unlabeled speech data and paired speech-text data. These two tasks are learnt to map speech into phonemes, which close the gap between speech and text and generate a good representation for ASR.

% MSP predicts phoneme distributions by masked language modeling with unlabeled speech data while PP employs CTC loss with paired speech-text data.

\subsubsection{Masked speech prediction}
\label{sec:MSP}
The MSP task utilizes unlabeled speech for self-supervised learning by masked language modeling, which has been proven good for pre-training speech encoders to generate a good representation of speech~\cite{baevski2020wav2vec,hsu2021hubert,baevski2022data2vec,tang2022unified}. Different targets are chosen for the masked frames in previous methods. In this paper, we choose the phoneme distributions as the predicted target like STPT~\cite{tang2022unified} since phonemes are a bridge between speech and text. On the other hand, ASR only cares about the pronunciation information in speech, ignoring additional other details, such as speaker, emotion, etc.
% The MSP task utilizes unlabeled speech for encoder pre-training by masked language modeling~\cite{devlin2018bert}. 
% Different targets are chosen for the masked frames in previous methods~\cite{baevski2020wav2vec,hsu2021hubert,baevski2022data2vec,tang2022unified}. 
% We choose the phoneme distributions as the predicted target~\cite{tang2022unified} for the masked frames since phonemes are a bridge between speech and text. 
Specifically, as shown in Fig.~\ref{model}, a speech feature extractor first extract latent speech representations from speech input $\mathbf{X}$, which can be represented as $\mathbf{Z}=(\mathbf{z}_1,\ldots,\mathbf{z}_{T})$.

Then the target and predicted phoneme distributions can be computed. Specifically, for the target phoneme distribution, the latent speech representations $Z$ are fed to the speech encoder and shared encoder to build contextualized speech representations $\mathbf{H}=\left(\mathbf{h_1},\ldots,\mathbf{h}_{T^{'}}\right)$, which are compared with a learnable phoneme embedding $\mathbf{E}=\left(\mathbf{e}_{1},\ldots,\mathbf{e}_{I}\right)$. $T^{'}$ is the down sampling size of $T$. Then the target phoneme distribution $p(\mathbf{e}_{i} | \mathbf{h}_{t})$ can be defined as follows: %\zc{incorrect expression. should be $p(e_i|h_t)$, at least the sum of p over i should be 1.}
\begin{equation}
    p(\mathbf{e}_{i} | \mathbf{h}_{t})  = \frac{\text{exp} (\mathbf{h}^{T}_{t} \cdot \mathbf{e}_{i}) }{\sum \limits_{j=1}^{I} \text{exp}(\mathbf{h}^{T}_{t} \cdot \mathbf{e}_{j}) }
\end{equation}
Note that the phoneme embedding $\mathbf{E}$ is actually frozen in the MSP task, which will be described in Section~\ref{sec:PP}. 

As for the predicted phoneme distribution, $\mathbf{\widetilde{Z}}$ is a masked version of $\mathbf{Z}$ obtaining by a span mask strategy\cite{baevski2022data2vec}. $\mathbf{\widetilde{H}}$ is the corresponding contextualized speech representations and $p( \mathbf{e}_{i} | \mathbf{\widetilde{h}}_{t})$ is the predicted phoneme distribution. Based on the target and predicted phoneme distributions, the masked KL divergence loss can be computed as follows:
\begin{equation}
    % \mathcal{L}_{\text{MSP}} = -\sum_{\widetilde{h}_t \in \mathcal{M}} 
    \mathcal{L}_{\text{MSP}} = -\sum_{t=1}^{T^{'}} 
    \sum \limits_{i=1}^{I} 
    p(\mathbf{e}_{i}| \mathbf{h}_{t} ) \text{log} \frac{p( \mathbf{e}_{i} |\mathbf{\widetilde{h}}_{t} )}{p(\mathbf{e}_{i} |\mathbf{h}_{t} )}
\end{equation}

\subsubsection{Phoneme prediction}
\label{sec:PP}
% A common issue of tasks which learn their own target is representation collapse. This occurs when the model predict very similar phoneme distribution for all masked frames resulting in a trivial task. 
% There exists representation collapse problem in the MSP task, namely ignoring the inputs when calculating the target and predicted phoneme distributions and produce identical and constant results, which makes the learning of the phoneme embedding essential. 
There exists the representation collapse problem in the MSP task since it learns its own generated target. This occurs when the model predicts similar phoneme distribution for all input masked frames resulting in a trivial task. Therefore, we introduce the PP task with paired speech-text data to guide phoneme embedding learning. Based on the predicted distribution $p(\mathbf{e}_{i} | \mathbf{h}_{t})$ and the target phoneme sequence converted from the corresponding text, we employ the CTC loss to optimize the PP task:
\begin{equation}
    \mathcal{L}_{\text{PP}} = -\sum_{\mathbf{e}_{t,j} \in \pi}\prod_{t=1}^{T'}p(\mathbf{e}_{t,j} | \mathbf{h}_{t})
\end{equation}
where $\pi$ denotes all possible augmented sequences with the blank symbol of the target phoneme sequence. 

To further alleviate the collapse problem, the phoneme embedding $\mathbf{E}$ is frozen in the MSP task and updated by the PP and P2T tasks. Besides, we find that spectrogram features are more stable than waveform features as inputs under the multi-task learning framework. The spectrogram features introduce fewer speech details, making it more difficult for the model to distinguish different task datasets. This way, the PP can prevent the MSP from being trivial effectively. 

% % 使用fbank的原因，kl_loss和监督任务一起训练，用wav特征会使得模型容易区分数据集。
% Besides, we find that spectrogram features are more stable than waveform features as inputs under the multi-task learning framework. The spectrogram features introduce fewer speech details, making it more difficult for the model to distinguish different task datasets. This way, the PP can avoid the MSP from trivial effectively. 

\subsection{Multi-task Pre-Training}
We additionally introduce the downstream speech-to-text task~(S2T) to further improve the pre-training performance:
\begin{equation}
\mathcal{L}_{\text{S2T}} = - \sum_{l=1}^{L} \text{log} p(\mathbf{y}^t_l|\mathbf{y}^t_{l-1},\mathbf{H})
\end{equation}
With the S2T task, we can actually obtain excellent recognition results even without fine-tuning.
Another advantage of introducing S2T is that we can directly estimate the quality of pre-training in the pre-training stage according to the results of S2T, which can facilitate the confirmation of pre-training related configuration.

Finally, the overall pre-training loss for MMSpeech can be defined as the weighted summation of five tasks losses:
\begin{equation}
    % \mathcal{L} = \lambda_{\text{MSP}} \mathcal{L}_{\text{MSP}} 
    % + \lambda_{\text{PP}} \mathcal{L}_{\text{PP}} 
    % + \lambda_{\text{S2C}} \mathcal{L}_{\text{S2C}} 
    % + \lambda_{\text{P2T}} \mathcal{L}_{\text{P2T}} 
    % + \lambda_{\text{S2T}} \mathcal{L}_{\text{S2T}}
    \mathcal{L} = \lambda_{1} \mathcal{L}_{\text{MSP}} 
    + \lambda_{2} \mathcal{L}_{\text{PP}} 
    + \lambda_{3} \mathcal{L}_{\text{S2C}} 
    + \lambda_{4} \mathcal{L}_{\text{P2T}} 
    + \lambda_{5} \mathcal{L}_{\text{S2T}}
\end{equation}
% where $\lambda_{1}=\lambda_{2} =\lambda_{3}=\lambda_{4}=\lambda_{5}=1$.
where $\lambda_{1}$, ... $\lambda_{5}$ are 
the task weights. 
% To stabilize the multi-task pre-training, we proposed to first train the model with the P2T task until convergence, and then conduct the multi-task pre-training base on it. We think there are two main reasons: 1) the P2T task can be seen as a simplified version of the S2T task, which lays the foundation for subsequent training; 2) the phoneme embedding $\mathbf{E}$ is well initialized by the P2T task before pre-training the MSP task, which prevent the encoder from collapsing. 

\section{Experiments}
\label{sec:exp}

\subsection{Experiment Settings}
\textbf{Data}
We evaluate MMSpeech on the Mandarin ASR corpus. For unlabeled speech data, AISHELL-2~\cite{du2018aishell} and WenetSpeech~\cite{zhang2022wenetspeech} corpus are separately considered. For unlabeled text data, we use M6-Corpus~\cite{lin2021m6} including 292GB texts. For paired speech-text data, we use AISHELL-1~\cite{bu2017aishell} corpus. All speech data is extracted to 80-dimensional log Mel-filterbank features while specAugment~\cite{park2019specaugment} and speed perturbation~\cite{ko2017study} are employed. A 21128-token BERT tokenizer\cite{devlin2018bert} is utilized to tokenize Mandarin texts.

\noindent
\textbf{Model configuration}
We implement MMSpeech based on OFA\footnote{\href{https://github.com/OFA-Sys/OFA}{https://github.com/OFA-Sys/OFA}. Our code shall be released soon.}~\cite{wang2022unifying}. We design MMSpeech in two different configurations, namely \textsc{Base} and \textsc{Large}. For \textsc{Base} architecture, the speech feature extractor is a two-layer convolutional network~(CNN) while each layer has 768 channels with strides (2,2) and kernel widths (3,3). The speech encoder, shared encoder and decoder all contain 6 transformer layers with model dimension 768, inner dimension 3,072 and 12 attention heads. For \textsc{Large} architecture, the transformer layer numbers of the speech encoder/shared encoder/decoder are changed to 12, the model dimensions are changed to 4096 and the attention heads are changed to 16, while other components are kept the same. 

\noindent
\textbf{Training detail}
\label{sec:train_detail}
To stabilize the multi-task pre-training, we proposed to first train the model with the P2T task until convergence and then conduct the multi-task pre-training based on it like~\cite{tang2022unified}. 
We think there are two main reasons: 
% 1) the P2T task can be seen as a simplified version of the S2T task, which lays the foundation for subsequent training; 
1) the P2T task initializes the phoneme embedding $\mathbf{E}$ well before pre-training the MSP task, which prevents the encoder from collapsing; 
2) the unlabeled text dataset utilized by the P2T task is much larger than the unlabeled speech and supervised speech-text dataset, making it hard to go through the whole text data during multi-task learning without overfitting other datasets.
For the S2C task, we use the Data2Vec model to generate pseudo-codes by default, which has a similar architecture with MMSpeech-\textsc{Base} encoder but with a 1D CNN~\cite{wang2020fairseqs2t} for feature extractor, and are pre-trained with 1000 hours AISHELL-2 audio. Besides, we set the number of K-means clusters to 500 and the vocabulary size of BPE to 30,000 by default.
For multi-task learning, we adjust the number of samples in a mini-batch to decide the task weights $\lambda_{1}$, \ldots $\lambda_{5}$. The sample rates in a mini-batch for each task are 4:4:2:1:1 for the MSP, S2C, P2T, PP, and S2T tasks, respectively. 
During pre-training, the mask ratio of phoneme inputs is set as 30\% for the P2T task. We select 7\% speech frames from the speech input as the mask span starting points with the mask span length of 10 for the MSP task. 
During inference, we use an external language model~(LM) for shallow fusion~\cite{gulcehre2015using} by default. The LM is implemented as a 12-layers Transformer and trained with the M6-Corpus text data with model dimension 768, inner dimension 3,072 and 12 attention heads. .

\subsection{Results}
We evaluate ASR performance with the word-error-rate~(WER) metric. Table \ref{main_reslut} presents AISHELL-1 recognition results of models under the \textsc{Base} configuration. For comparison, we also evaluate the pre-trained Data2Vec mentioned in Section~\ref{sec:train_detail}. We fine-tune the Data2Vec with the same supervised data as MMSpeech and a CTC loss. The speech data used for all pre-training methods is AISHELL-2 audio. As shown in Table \ref{main_reslut}, our proposed MMSpeech outperforms the model without pre-training and the Data2Vec significantly, whether with or without LM fusion. Furthermore, MMSpeech can obtain excellent recognition results even without fine-tuning (FT), as shown in the fourth row in Table \ref{main_reslut}. 
% The following experiments in Section~\ref{sec:ab} are based on this model.

\begin{table}[htb]
    \setlength\tabcolsep{4pt}
	\centering
	\caption{WER on the AISHELL-1 dev/test set when the unlabeled speech data is AISHELL-2 audio and the architecture is the \textsc{Base}.}
    \vspace{0.1cm}
    \label{main_reslut}
	\begin{tabular}{l c c c c c}
		\toprule  % 顶部线
		\multirow{2}{*}{Model} 
		& \multicolumn{2}{c}{dev} & \multicolumn{2}{c}{test} \\
		\cmidrule(lr){2-3} \cmidrule(lr){4-5}
         & w/o LM & with LM & w/o LM & with LM \\
		\midrule  % 中部线
        w/o pre-training     & 6.4   & 5.2   & 6.8   & 5.7 \\
        Data2Vec            & 3.8   & 3.7   & 4.1   & 3.9 \\
        % MMSpeech            & 2.5   & 2.3   & 2.6   & 2.3 \\
        \textbf{MMSpeech}   & \textbf{2.4} & \textbf{2.1} &\textbf{2.6} & \textbf{2.3} \\
        ~--~~w/o FT           & 2.5   & 2.3   & 2.6   & 2.3 \\
		\bottomrule  % 底部线
	\end{tabular}
\end{table}

We also compare our methods with the previously published Mandarin pre-training models\footnote{\href{https://github.com/TencentGameMate/chinese\_speech\_pretrain}{https://github.com/TencentGameMate/chinese\_speech\_pretrain}} in Table~\ref{compare_with_published}, where the Wav2Vec 2.0\cite{baevski2020wav2vec} and HuBERT\cite{hsu2021hubert} are ASR systems which are pre-trained with WenetSpeech audio and fine-tuned on AISHELL-1. 
We use the same speech data WenetSpeech to pre-train MMSpeech. 
Besides, we use the above pre-trained HuBERT-\textsc{Base} to generate pseudo codes for the S2C task.
% We compare models under the same unlabeled speech data and encoder size.
Table \ref{compare_with_published} shows the WER results rescoring with an LM. We achieve improvement from Table~\ref{main_reslut} since the WenetSpeech dataset contains ten times the audio data of AISHELL-2. Furthermore, our proposed MMSpeech outperforms Wav2Vec 2.0\cite{baevski2020wav2vec} and HuBERT\cite{hsu2021hubert} under the \textsc{Base} or \textsc{Large} setting. We obtain the SOTA performance on the AISHELL-1 dev/test set, achieving a relative 48.3\%/42.4\% WER decrease. 

% \begin{table}[htb]
%     \vspace{-1.0em}
%     \setlength\tabcolsep{4pt}
% 	\centering
% 	\caption{WER on the AISHELL-1 dev/test set when the unlabeled speech data is AISHELL-2 audio and the architecture is the \textsc{Base}.}
%     \label{main_reslut}
% 	\begin{tabular}{l c c c c c}
% 		\toprule  % 顶部线
% 		Model & dev & test \\
% 		\midrule  % 中部线
%         w/o pre-training     & 6.4 (5.2)   & 6.8 (5.7) \\
%         Data2Vec            & 3.8 (3.7)   & 4.1 (3.9) \\
%         % MMSpeech            & 2.5   & 2.3   & 2.6   & 2.3 \\
%         \textbf{MMSpeech}   & \textbf{2.4} (\textbf{2.1}) &\textbf{2.6} (\textbf{2.3}) \\
%         ~--~~w/o FT           & 2.5 (2.3)   & 2.6 (2.3) \\
% 		\bottomrule  % 底部线
% 	\end{tabular}
%     \vspace{-1.0em}
% \end{table}

\begin{table}[htb]
	\centering
	\caption{WER on the AISHELL-1 dev/test sets compared with the published models pre-trained with the WenetSpeech audio.}
    \vspace{0.1cm}
    \label{compare_with_published}
	\begin{tabular}{l c c c c c}
		\toprule
		Model & encoder size & dev & test \\ 
		\midrule
        Wav2Vec 2.0         & \textsc{Base}   & 4.2   & 4.7 \\
        HuBERT              & \textsc{Base}    & 4.1   & 4.3 \\
        \textbf{MMSpeech}   & \textsc{Base}    & \textbf{2.0}  & \textbf{2.1} \\
        \midrule
        Wav2Vec 2.0         & \textsc{Large}   & 3.8   & 4.1 \\
        HuBERT              & \textsc{Large}   & 3.1   & 3.3 \\
        \textbf{MMSpeech}   & \textsc{Large}   & \textbf{1.6}  & \textbf{1.9} \\
		\bottomrule
	\end{tabular}
\end{table}

\subsection{Ablation study}
\label{sec:ab}
To better understand MMSpeech, we conduct an ablation study by removing different pre-training tasks for multi-task learning. We use the same configuration as Table~\ref{main_reslut} and present the result on the AISHELL-1 dev/test set. As shown in Table \ref{ablation_study}, we have the following findings: (1) In the second row, we observe significant performance degradation when pre-training without the P2T task. Compared with the fifth row, unlabeled text data plays a more critical role than unlabeled speech data in MMSpeech. It is different from the conclusion of SpeechT5\cite{ao2021speecht5}, since our unlabeled text data is huge, 292GB, while only 1.8GB of texts are used previously\cite{ao2021speecht5,tang2022unified}. Even decoding with an LM, introducing the P2T task still contributes an average 0.65 WER reduction, which is different from STPT~\cite{tang2022unified} and proves that the Mandarin P2T task not only learns linguistic information from text data but also acts as a supplement to the S2T modeling. (2) In the third to the fifth row, we present the results without the unlabeled speech tasks. The S2C and MSP tasks are important to MMSpeech since the performance degrades significantly without each of them. One interesting observation is that the model pre-trained without unlabeled speech data outperforms the model pre-trained with an additional S2C task when decoding without LM, as shown in the fifth and third rows. It demonstrates that the encoder pre-training tasks are essential. The S2C task should be optimized with the MSP task jointly, where the MSP task helps learn better speech representation. (3) In the sixth row, we remove the PP task for the joint pre-training, which makes the phoneme embedding learning without guidance and causes significant WER increase. The PP task can benefit the pre-training by aligning the phoneme representation and speech representation. Furthermore, the training doesn't converge when removing the PP and S2C tasks together, and we observe that all predictions in the MSP collapse into one or two target phonemes. It proves that the MSP task can easily cause the representation collapse problem while multi-task learning prevents it. (4) In the last row, we remove the supervised S2T task during pre-training but keep the same number of training steps. The WER has an average 0.6 increase after fine-tuning, proving that introducing the downstream S2T task can speed up the model convergence on the downstream tasks and improve performance. Besides, we can directly estimate the quality of pre-training in the pre-training stage according to the results of S2T, which can facilitate the confirmation of pre-training-related configuration.
% 1) All the pre-training tasks are essential for the MMSpeech. 2) Different from the conclusion of SpeechT5\cite{ao2021speecht5} and STPT\cite{tang2022unified}, the unsupervised text task P2T plays the most critical role in multi-task learning. It's because the amount of our unlabeled text data is huge, 292GB, while only 1.8GB of texts are used previously\cite{ao2021speecht5,tang2022unified}. 3) In the fifth row, the model is trained without unlabeled speech data, i.e., only the MPP, P2T, and S2T tasks are optimized, which outperforms the model that MPP, P2T, S2T, and S2C are optimized as shown in the third row. It demonstrates that the S2C task should be optimized with the MSP task jointly, where the MSP task acts as a regularizer for the S2C task. 4) In the last row, we remove the supervised S2T task during pre-training but keep the same number of training steps. The WER has avarage ? increase after fine-tuning.
\begin{table}[htb]
	\centering
	\caption{Ablation study based on MMSpeech in Table~\ref{main_reslut}.}
    \label{ablation_study}
	\begin{tabular}{l c c c c c}
		\toprule
		& \multicolumn{2}{c}{dev} & \multicolumn{2}{c}{test} \\
        \cmidrule(lr){2-3} \cmidrule(lr){4-5}
         & w/o LM & with LM & w/o LM & with LM \\
		\midrule
        \textbf{MMSpeech}        & \textbf{2.4}   & \textbf{2.1}   & \textbf{2.6} & \textbf{2.3}\\
        --~~P2T         & 3.4   & 2.7   & 3.8 & 3.0 \\
        --~~MSP         & 2.9   & 2.4   & 3.2 & 2.6\\
        --~~S2C         & 2.6   & 2.3   & 2.8 & 2.5 \\
        --~~MSP\&S2C    & 2.7   & 2.4  & 3.1 & 2.7 \\
        --~~PP          & 3.1   & 2.5  & 3.5 & 2.8 \\
        --~~S2T         & 2.9   & 2.4  & 3.3 & 2.7 \\
		\bottomrule
	\end{tabular}
\end{table}

\subsection{Analysis}
\label{sec:an}
\subsubsection{Impact of unsupervised text task} 
To analyze why the P2T task is effective, we investigate two main factors: input features and text data amount for the P2T task. Firstly, as shown in the second and third row in Table~\ref{unsup_text}, we compare the recognition results from jointly pre-training the supervised tasks and P2T task with that from jointly pre-training the supervised tasks and text-infilling~(T2T) task. The P2T task significantly outperforms the T2T task as phoneme bridges the gap between Mandarin text and speech, making it easier for the speech and text inputs to share an encoder. Besides, with an external LM, the P2T task obtains more obvious performance improvements than the T2T task, that an average of 0.33 WER reduction observed for P2T and an average of 0.17 WER reduction for T2T. It is because the P2T task learns not only the language modeling abilities but also the relationship between pronunciation and text, which is important for Mandarin ASR since Mandarin has a large number of homophones. 

Secondly, in the fourth row, we sample 1.8G text data from the M6-corpus to obtain the same amount of text data as the previous works~\cite{ao2021speecht5, tang2022unified}. We observe a performance degradation of the P2T task when reducing the text data while it still outperforms the T2T task utilizing larger text data. Furthermore, unlike STPT~\cite{tang2022unified}, which conducts experiments on the English dataset, our experiments show the improvement of the P2T task can not be replaced by an LM when using the same amount of text data. It indicates that the P2T task is more effective than Mandarin tasks due to the larger difference between Mandarin speech and text.

\begin{table}[htb]
	\centering
	\caption{Comparision of the unsupervised text tasks on AISHELL-1 dev/test set. “()” indicates the WER is measured with an external LM.}
    \vspace{0.1cm}
    \label{unsup_text}
	\begin{tabular}{lccccc}
		\toprule  % 顶部线
		Model & unsup-text & dev & test \\
		\midrule  % 中部线
        w/o pre-training & - & 6.4 (5.2) & 6.8 (5.7) \\
        + P2T         & 292G & 2.7 (2.4) & 3.1 (2.7) \\
        + T2T         & 292G & 3.7 (3.6) & 4.2 (3.9) \\
        + P2T         & 1.8G & 3.0 (2.8)   & 3.5 (3.2) \\
		\bottomrule  % 底部线
	\end{tabular}
\end{table}

\subsubsection{Generalizability of pretrained models} 
We evaluate the AISEHLL-2 test set to validate the generalizability of MMSpeech. Table~\ref{aishell2} reports the recognition results with an external LM. The first part of the table shows the published results, which are from the ASR models trained with AISHELL-2 supervised data without pre-training. Transformer is an ASR model with a similar structure to MMSpeech and Conformer is the SOTA model for AISHELL-2. 

The second part of Table~\ref{aishell2} presents results from the MMSpeech-\textsc{Base}, where the pre-trained MMSpeech model is the same as that in Table~\ref{main_reslut}. Differently, we fine-tune the MMSpeech-\textsc{Base} model with the AISHELL-2 train set. MMSpeech-\textsc{Base} outperforms the published SOTA results on the iOS and Android test sets. Surprisingly, the model performs the worst on the Microphone test set, while the labeled data AISHELL-1 used by the S2T task during pre-training is exactly collected with the Microphone. We consider it because the collection methods of the labeled and unlabeled speech data are single in their own set but significantly different from each other, which makes the pre-trained model less robust. Specifically, the labeled speech data AISHELL-1 is collected via the Microphone device, and the unlabeled speech data AISHELL-2 is collected via the iPhone device. They are collected from the different relative positions of a speaker as described in~\cite{du2018aishell}. Under multi-task learning, these different task datasets can be easily distinguished, which dramatically reduces the effect of joint training. Furthermore, the problem gets worse when replacing the log Mel-filterbank feature with the raw audio waveform as the speech input, as shown in the third row. The model is even invalid on the iOS/Android test set. We suppose that waveform features contain more speech detail, which makes the model easier to distinguish different task datasets and causes the multi-task learning invalid.

In the third part of Table~\ref{aishell2}, we utilize the WenetSpeech~\cite{zhang2022wenetspeech} audio as the unlabeled speech data for pre-training, which has diverse speech data collected from the Internet with multiple speaking, styles, scenarios, domains, topics, and noisy conditions. The pre-trained model is much more robust. As shown in the second row, even without fine-tuning and using only AISHELL-1~(178h) supervised data during pre-training, MMSpeech can achieve comparable results with the model trained with AISHELL-2~(1000h). Besides, MMSpeech, after fine-tuning with the AISHELL-2 train set, outperforms the published SOTA results on the AISHELL-2 test set. It proves that the diversity of pre-training data is important. In the future, we will replace the supervised data AISHELL-1 with more diverse data during pre-training expecting to obtain better generalizability of the pre-trained model.

% We compare our results to the ASR models trained with AISHELL-2 supervised data without pre-training. 
% Table~\ref{aishell2} shows that our model without fine-tuning and using only AISHELL-1~(178h) supervised data during pre-training can achieve a comparable results with the model trained with AISHELL-2~(1000h) in the fourth row. Besides, MMSpeech after fine-tuning with the AISHELL-2 train set outperforms the published SOTA results on the AISHELL-2 test set.
% We also find that the pre-trained model using waveform features as input can not converage on AISHELL-2 iOS and Android test set. It's because AISHELL-1 is collected by Mic, while AISHELL-2 train set is collected by iOS. They are employed for MSP and PP tasks, respectively. Waveform features is easy to distinguish different task datasets, which make the MSP lead to representation collapse.
% wav特征和fbank特征的选择

\begin{table}[htb]
    \setlength\tabcolsep{3pt}
	\centering
	\caption{MMSpeech-\textsc{Base} evaluated on AISHELL-2 test set.}
    \label{aishell2}
	\begin{tabular}{lccccc}
		\toprule
		Model   & unsup-speech  & iOS  & Mic  & Android \\ 
		\midrule
        Transformer\tablefootnote{Transformer results from https://github.com/espnet/espnet}~\cite{mohamed2019transformers}      &-     & 7.5   & 8.6   & 8.9 \\
        Conformer\tablefootnote{SOTA results from https://paperswithcode.com/}~\cite{gulati2020conformer}   &-     & 5.3   & 5.6   & 5.7 \\
        \midrule
        MMSpeech    & AISHELL-2  &   3.9 &   7.4     &    4.1 \\
        - w/o FT    & AISHELL-2  &   6.7 &   10.4    &   6.6 \\
        $\xrightarrow{}$ wav(w/o FT)    &   AISHELL-2   &   -   &   15.3    &   - \\
        \midrule
        \textbf{MMSpeech}    &   WenetSpeech &   \textbf{3.9}    &   \textbf{4.5}    &   \textbf{4.0} \\
        - w/o FT    &   WenetSpeech &   6.2 &   7.1 &   6.5 \\
		\bottomrule
	\end{tabular}
\end{table}
\section{Conclusion}
MMSpeech employs a multi-task learning framework with five self-supervised and supervised tasks. Unlabeled speech and text data are used to pre-train both the encoder and decoder and are built a relation with phonemes. By introducing the phoneme modality into pre-training, we build a bridge between speech and text, which is especially beneficial for Mandarin ASR. Experiments show MMSpeech obtain the SOTA performance on the AISHELL-1. Furthermore, we conduct a detailed ablation study and analysis to understand the five pre-training tasks. The proposed five tasks are important to MMSpeech. Besides, we find the P2T task utilizing unlabeled text data achieves the most significant improvement and can not be replaced by an external LM. Therefore, we reaffirm the importance of unlabeled text for ASR pre-training.

\bibliographystyle{IEEEbib}
\bibliography{main.bib}

\begin{thebibliography}{10}

\bibitem{schneider2019wav2vec}
Steffen Schneider, Alexei Baevski, Ronan Collobert, and Michael Auli,
\newblock ``wav2vec: Unsupervised pre-training for speech recognition,''
\newblock {\em arXiv preprint arXiv:1904.05862}, 2019.

\bibitem{baevski2020wav2vec}
Alexei Baevski, Yuhao Zhou, Abdelrahman Mohamed, and Michael Auli,
\newblock ``wav2vec 2.0: A framework for self-supervised learning of speech
  representations,''
\newblock {\em Advances in Neural Information Processing Systems}, vol. 33, pp.
  12449--12460, 2020.

\bibitem{hsu2021hubert}
Wei-Ning Hsu, Benjamin Bolte, Yao-Hung~Hubert Tsai, Kushal Lakhotia, Ruslan
  Salakhutdinov, and Abdelrahman Mohamed,
\newblock ``Hubert: Self-supervised speech representation learning by masked
  prediction of hidden units,''
\newblock {\em IEEE/ACM Transactions on Audio, Speech, and Language
  Processing}, vol. 29, pp. 3451--3460, 2021.

\bibitem{baevski2022data2vec}
Alexei Baevski, Wei-Ning Hsu, Qiantong Xu, Arun Babu, Jiatao Gu, and Michael
  Auli,
\newblock ``Data2vec: A general framework for self-supervised learning in
  speech, vision and language,''
\newblock {\em arXiv preprint arXiv:2202.03555}, 2022.

\bibitem{graves2006connectionist}
Alex Graves, Santiago Fern{\'a}ndez, Faustino Gomez, and J{\"u}rgen
  Schmidhuber,
\newblock ``Connectionist temporal classification: labelling unsegmented
  sequence data with recurrent neural networks,''
\newblock in {\em Proceedings of the 23rd international conference on Machine
  learning}, 2006, pp. 369--376.

\bibitem{chan2015listen}
William Chan, Navdeep Jaitly, Quoc~V Le, and Oriol Vinyals,
\newblock ``Listen, attend and spell,''
\newblock {\em arXiv preprint arXiv:1508.01211}, 2015.

\bibitem{mohamed2019transformers}
Abdelrahman Mohamed, Dmytro Okhonko, and Luke Zettlemoyer,
\newblock ``Transformers with convolutional context for asr,''
\newblock {\em arXiv preprint arXiv:1904.11660}, 2019.

\bibitem{ao2022pre}
Junyi Ao, Ziqiang Zhang, Long Zhou, Shujie Liu, Haizhou Li, Tom Ko, Lirong Dai,
  Jinyu Li, Yao Qian, and Furu Wei,
\newblock ``Pre-training transformer decoder for end-to-end asr model with
  unpaired speech data,''
\newblock {\em arXiv preprint arXiv:2203.17113}, 2022.

\bibitem{ao2021speecht5}
Junyi Ao, Rui Wang, Long Zhou, Shujie Liu, Shuo Ren, Yu~Wu, Tom Ko, Qing Li,
  Yu~Zhang, Zhihua Wei, et~al.,
\newblock ``Speecht5: Unified-modal encoder-decoder pre-training for spoken
  language processing,''
\newblock {\em arXiv preprint arXiv:2110.07205}, 2021.

\bibitem{tang2022unified}
Yun Tang, Hongyu Gong, Ning Dong, Changhan Wang, Wei-Ning Hsu, Jiatao Gu,
  Alexei Baevski, Xian Li, Abdelrahman Mohamed, Michael Auli, et~al.,
\newblock ``Unified speech-text pre-training for speech translation and
  recognition,''
\newblock {\em arXiv preprint arXiv:2204.05409}, 2022.

\bibitem{zhou2015homophone}
Wei Zhou,
\newblock {\em The Homophone Effect in Mandarin Word Recognition},
\newblock Ph.D. thesis, The Ohio State University, 2015.

\bibitem{lin2021m6}
Junyang Lin, Rui Men, An~Yang, Chang Zhou, Ming Ding, Yichang Zhang, Peng Wang,
  Ang Wang, Le~Jiang, Xianyan Jia, et~al.,
\newblock ``M6: A chinese multimodal pretrainer,''
\newblock {\em arXiv preprint arXiv:2103.00823}, 2021.

\bibitem{Tang2021AGM}
Yun Tang, J.~Pino, Changhan Wang, Xutai Ma, and Dmitriy Genzel,
\newblock ``A general multi-task learning framework to leverage text data for
  speech to text tasks,''
\newblock in {\em ICASSP}, 2021.

\bibitem{lewis2019bart}
Mike Lewis, Yinhan Liu, Naman Goyal, Marjan Ghazvininejad, Abdelrahman Mohamed,
  Omer Levy, Ves Stoyanov, and Luke Zettlemoyer,
\newblock ``Bart: Denoising sequence-to-sequence pre-training for natural
  language generation, translation, and comprehension,''
\newblock {\em arXiv preprint arXiv:1910.13461}, 2019.

\bibitem{oord2018representation}
Aaron van~den Oord, Yazhe Li, and Oriol Vinyals,
\newblock ``Representation learning with contrastive predictive coding,''
\newblock {\em arXiv preprint arXiv:1807.03748}, 2018.

\bibitem{devlin2018bert}
Jacob Devlin, Ming-Wei Chang, Kenton Lee, and Kristina Toutanova,
\newblock ``Bert: Pre-training of deep bidirectional transformers for language
  understanding,''
\newblock {\em arXiv preprint arXiv:1810.04805}, 2018.

\bibitem{chen2022wavlm}
Sanyuan Chen, Chengyi Wang, Zhengyang Chen, Yu~Wu, Shujie Liu, Zhuo Chen, Jinyu
  Li, Naoyuki Kanda, Takuya Yoshioka, Xiong Xiao, et~al.,
\newblock ``Wavlm: Large-scale self-supervised pre-training for full stack
  speech processing,''
\newblock {\em IEEE Journal of Selected Topics in Signal Processing}, 2022.

\bibitem{wu2022wav2seq}
Felix Wu, Kwangyoun Kim, Shinji Watanabe, Kyu Han, Ryan McDonald, Kilian~Q
  Weinberger, and Yoav Artzi,
\newblock ``Wav2seq: Pre-training speech-to-text encoder-decoder models using
  pseudo languages,''
\newblock {\em arXiv preprint arXiv:2205.01086}, 2022.

\bibitem{lu2019vilbert}
Jiasen Lu, Dhruv Batra, Devi Parikh, and Stefan Lee,
\newblock ``Vilbert: Pretraining task-agnostic visiolinguistic representations
  for vision-and-language tasks,''
\newblock {\em Advances in neural information processing systems}, vol. 32,
  2019.

\bibitem{radford2021learning}
Alec Radford, Jong~Wook Kim, Chris Hallacy, Aditya Ramesh, Gabriel Goh,
  Sandhini Agarwal, Girish Sastry, Amanda Askell, Pamela Mishkin, Jack Clark,
  et~al.,
\newblock ``Learning transferable visual models from natural language
  supervision,''
\newblock in {\em International Conference on Machine Learning}. PMLR, 2021,
  pp. 8748--8763.

\bibitem{wang2022unifying}
Peng Wang, An~Yang, Rui Men, Junyang Lin, Shuai Bai, Zhikang Li, Jianxin Ma,
  Chang Zhou, Jingren Zhou, and Hongxia Yang,
\newblock ``Unifying architectures, tasks, and modalities through a simple
  sequence-to-sequence learning framework,''
\newblock {\em arXiv preprint arXiv:2202.03052}, 2022.

\bibitem{bapna2021slam}
Ankur Bapna, Yu-an Chung, Nan Wu, Anmol Gulati, Ye~Jia, Jonathan~H Clark,
  Melvin Johnson, Jason Riesa, Alexis Conneau, and Yu~Zhang,
\newblock ``Slam: A unified encoder for speech and language modeling via
  speech-text joint pre-training,''
\newblock {\em arXiv preprint arXiv:2110.10329}, 2021.

\bibitem{han2021learning}
Chi Han, Mingxuan Wang, Heng Ji, and Lei Li,
\newblock ``Learning shared semantic space for speech-to-text translation,''
\newblock {\em arXiv preprint arXiv:2105.03095}, 2021.

\bibitem{Tang2021IST}
Yun Tang, Juan Pino, Xian Li, Changhan Wang, and Dmitriy Genzel,
\newblock ``Improving speech translation by understanding and learning from the
  auxiliary text translation task,''
\newblock in {\em ACL}, 2021.

\bibitem{bapna2022mslam}
Ankur Bapna, Colin Cherry, Yu~Zhang, Ye~Jia, Melvin Johnson, Yong Cheng, Simran
  Khanuja, Jason Riesa, and Alexis Conneau,
\newblock ``mslam: Massively multilingual joint pre-training for speech and
  text,''
\newblock {\em arXiv preprint arXiv:2202.01374}, 2022.

\bibitem{vaswani2017attention}
Ashish Vaswani, Noam Shazeer, Niki Parmar, Jakob Uszkoreit, Llion Jones,
  Aidan~N Gomez, {\L}ukasz Kaiser, and Illia Polosukhin,
\newblock ``Attention is all you need,''
\newblock {\em Advances in neural information processing systems}, vol. 30,
  2017.

\bibitem{gage1994new}
Philip Gage,
\newblock ``A new algorithm for data compression,''
\newblock {\em C Users Journal}, vol. 12, no. 2, pp. 23--38, 1994.

\bibitem{du2018aishell}
Jiayu Du, Xingyu Na, Xuechen Liu, and Hui Bu,
\newblock ``Aishell-2: Transforming mandarin asr research into industrial
  scale,''
\newblock {\em arXiv preprint arXiv:1808.10583}, 2018.

\bibitem{zhang2022wenetspeech}
Binbin Zhang, Hang Lv, Pengcheng Guo, Qijie Shao, Chao Yang, Lei Xie, Xin Xu,
  Hui Bu, Xiaoyu Chen, Chenchen Zeng, et~al.,
\newblock ``Wenetspeech: A 10000+ hours multi-domain mandarin corpus for speech
  recognition,''
\newblock in {\em ICASSP 2022-2022 IEEE International Conference on Acoustics,
  Speech and Signal Processing (ICASSP)}. IEEE, 2022, pp. 6182--6186.

\bibitem{bu2017aishell}
Hui Bu, Jiayu Du, Xingyu Na, Bengu Wu, and Hao Zheng,
\newblock ``Aishell-1: An open-source mandarin speech corpus and a speech
  recognition baseline,''
\newblock in {\em 2017 20th conference of the oriental chapter of the
  international coordinating committee on speech databases and speech I/O
  systems and assessment (O-COCOSDA)}. IEEE, 2017, pp. 1--5.

\bibitem{park2019specaugment}
Daniel~S Park, William Chan, Yu~Zhang, Chung-Cheng Chiu, Barret Zoph, Ekin~D
  Cubuk, and Quoc~V Le,
\newblock ``Specaugment: A simple data augmentation method for automatic speech
  recognition,''
\newblock {\em arXiv preprint arXiv:1904.08779}, 2019.

\bibitem{ko2017study}
Tom Ko, Vijayaditya Peddinti, Daniel Povey, Michael~L Seltzer, and Sanjeev
  Khudanpur,
\newblock ``A study on data augmentation of reverberant speech for robust
  speech recognition,''
\newblock in {\em 2017 IEEE International Conference on Acoustics, Speech and
  Signal Processing (ICASSP)}. IEEE, 2017, pp. 5220--5224.

\bibitem{wang2020fairseqs2t}
Changhan Wang, Yun Tang, Xutai Ma, Anne Wu, Dmytro Okhonko, and Juan Pino,
\newblock ``fairseq s2t: Fast speech-to-text modeling with fairseq,''
\newblock in {\em Proceedings of the 2020 Conference of the Asian Chapter of
  the Association for Computational Linguistics (AACL): System Demonstrations},
  2020.

\bibitem{gulcehre2015using}
Caglar Gulcehre, Orhan Firat, Kelvin Xu, Kyunghyun Cho, Loic Barrault, Huei-Chi
  Lin, Fethi Bougares, Holger Schwenk, and Yoshua Bengio,
\newblock ``On using monolingual corpora in neural machine translation,''
\newblock {\em arXiv preprint arXiv:1503.03535}, 2015.

\bibitem{gulati2020conformer}
Anmol Gulati, James Qin, Chung-Cheng Chiu, Niki Parmar, Yu~Zhang, Jiahui Yu,
  Wei Han, Shibo Wang, Zhengdong Zhang, Yonghui Wu, et~al.,
\newblock ``Conformer: Convolution-augmented transformer for speech
  recognition,''
\newblock {\em arXiv preprint arXiv:2005.08100}, 2020.

\end{thebibliography}

\end{document}